\documentclass[a4paper,11pt]{article}
\usepackage{jheppub} 

\usepackage[T1]{fontenc}
\usepackage{amsthm,amsmath,amssymb}
\usepackage{mathrsfs}
\usepackage{caption}
\usepackage{subfigure}
\usepackage{graphicx}
\usepackage{bm}
\usepackage{amsmath}
\usepackage{amsfonts}
\usepackage{amssymb}
\usepackage{bbm}
\usepackage[utf8]{inputenc}

\title{Gravitational form factors of the baryon octet in holographic QCD}

\author[a]{Zhibo Liu,}
\author[b]{Hiroaki Nakajima,}
\author[c]{Hiroaki Abuki,}
\author[d]{and Akira Watanabe}
\affiliation[a]{Department of Physics, Nagoya University, \\ Nagoya 464-8602, Japan}
\affiliation[b]{School of Mathematics and Physics, University of South China, \\ Hengyang 421001, People’s Republic of China}
\affiliation[c]{ Department of Education, Aichi University of Education, \\ 1 Hirosawa, Kariya 448-8542, Japan}
\affiliation[d]{National Institute of Technology, Oshima College, \\ Oshima 742-2193, Japan}

\emailAdd{zhibo.liu.hep@gmail.com}
\emailAdd{nakajima31158@gmail.com}
\emailAdd{abuki@auecc.aichi-edu.ac.jp}
\emailAdd{watanabe.akira@oshima-k.ac.jp}

\abstract{
The gravitational form factors (GFFs) of the baryon octet, including hyperons, are
investigated in a bottom-up holographic QCD model that explicitly incorporates the SU(3)
flavor symmetry breaking through the strange quark mass.
We fit the model parameters to reproduce the empirical masses of the baryon octet and
examine the dependence of GFFs on the probe momentum.
Our numerical results show distinct differences in the GFFs across the baryon octet.
The computed GFFs are found to be in reasonable agreement with available lattice QCD results for the non-strange/nucleon sector.
We also calculate the gravitational radii of the baryon octet and find that they
decrease with increasing strangeness, indicating that heavier hyperons are more compact.
}

\begin{document}
\maketitle
\flushbottom

\section{Introduction}
\label{sec:introduction}

Quantum chromodynamics (QCD), the fundamental theory describing the strong interaction
within the framework of the standard model, is a non-Abelian gauge theory based on the color
symmetry group SU(3) \cite{Gross:1973id, Politzer:1973fx}. 
It successfully accounts for the rich phenomenology of hadrons through the dynamics
of quarks and gluons.
Perturbative QCD is applicable in the high-energy regime, where the asymptotic freedom
ensures that the coupling becomes sufficiently weak to allow the perturbative
expansions~\cite{Lepage:1980fj, Tanaka:2018wea, Tong:2022zax}. 
In contrast, at low energies, the coupling becomes large, rendering perturbative
techniques ineffective. 
In this regime, essential features of hadron physics, such as the color confinement,
spontaneous chiral symmetry breaking, mass generation,
and formation of complex bound states, must be approached via effective or
non-perturbative methods~\cite{Callan:1977gz,Hutauruk:2016sug, daSilva:2012gf, Raya:2021zrz}.

One particularly powerful but underutilized probe for the internal structure of hadrons
is the energy-momentum tensor (EMT).
The matrix elements of EMT between the hadronic states encode the fundamental information
about the spatial distributions of the energy, momentum, and mechanical stress inside the
hadrons.
These matrix elements define a set of Lorentz-invariant gravitational form factors (GFFs),
which provide a field-theoretically robust framework for describing the hadron mechanics.
Although direct gravitational probes are experimentally infeasible at hadronic scales,
GFFs can be accessed indirectly through the study of generalized parton distributions
(GPDs)~\cite{Belitsky:2005qn,Belitsky:2001ns,Diehl:2003ny}, which appear in the deeply
virtual Compton scattering~\cite{Ji:1996nm}, deeply virtual meson production, and other
exclusive processes.

In recent years, interest in GFFs has been steadily growing.
Foundational works have clarified their connection to the EMT and established a
framework for the mechanical properties of hadrons~%
\cite{Polyakov:2002yz,Belitsky:2002jp,Kumano:2017lhr,Polyakov:2018exb,Polyakov:2019lbq,Metz:2021lqv}.
GFFs have also been obtained in lattice QCD~%
\cite{Shanahan:2018pib,Alexandrou:2021ztx, Hackett:2023rif,Hackett:2023nkr}. 
Model-based approaches have been widely used to study GFFs.
These include holographic QCD~\cite{Mamo:2022eui,Abidin:2008hn,Abidin:2008ku,Abidin:2009hr, Liu:2025vfe},
light-front QCD~\cite{Brodsky:2008pf, Chakrabarti:2015lba},
QCD sum rules~\cite{Anikin:2019kwi, Aliev:2020aih, Ozdem:2020ieh},
NJL model~\cite{Freese:2019eww,Freese:2019bhb},
and other effective models~%
\cite{Broniowski:2008hx, Shi:2020pqe,Kim:2020lrs,Krutov:2020ewr,Tong:2021ctu,Krutov:2022zgg,%
Guo:2023pqw,Guo:2023qgu,Xu:2023izo,Hatta:2023fqc,Won:2022cyy}.
Despite these efforts, studies of baryons beyond the nucleon remain limited, especially
in the strange sector.

Hyperons, which contain one or more strange quarks, constitute an essential part of the
low-lying baryon spectra.
As members of the flavor SU(3) baryon octet, the hyperons such as $\Lambda$, $\Sigma$,
and $\Xi$ provide a natural laboratory for studying the explicit flavor symmetry breaking,
the effect of strangeness on hadronic structure, and the interplay between constituent
quark masses and gluonic dynamics~\cite{Jenkins:1992se}.
They exhibit well-characterized weak decay modes that can, in principle, be exploited
for structural studies.
Furthermore, understanding the internal structure of hyperons is relevant in nuclear
astrophysics, where the hyperon degrees of freedom influence the equation of state of
dense matter in neutron stars.

Theoretically, GFFs of hyperons illuminate how the presence of strange quarks
alters the spatial structure of the EMT, including energy-momentum densities and
the associated mechanical properties.
In particular, the mass radius extracted from the hyperon GFF \(A(Q^2)\) is
different from that of the nucleon, reflecting the role of the strange-quark mass
and condensate in shaping internal dynamics.
Understanding the SU(3) flavor symmetry breaking in these distributions remains an
open problem and motivates us to develop a common, predictive framework for nucleons
and hyperons~\cite{Won:2022cyy,Petschauer:2020urh,Ozdem:2020ieh}.

The AdS/CFT correspondence~\cite{Veneziano:1968yb, Maldacena:1997re, Gubser:1998bc, Witten:1998qj},
also known as gauge/gravity duality, offers a novel and powerful approach to the study of strongly
coupled gauge theories.
In this framework, a strongly interacting four-dimensional superconformal field theory is mapped
to a weakly coupled five-dimensional supergravity in the AdS space.
Although QCD is neither conformal nor supersymmetric, bottom-up holographic QCD models have been
developed to incorporate key features of QCD by deforming the AdS background geometry~%
\cite{Sakai:2004cn, Sakai:2005yt, Kruczenski:2003uq, Aharony:1999ti, Erlich:2005qh}.
This deformation includes the introduction of an infrared (IR) cutoff (the hard-wall model~%
\cite{Erlich:2005qh}) or of a background dilaton field (the soft-wall model~\cite{Karch:2006pv})
to mimic the confinement and reproduce the Regge behavior.

In the context of hadron physics, bottom-up holographic QCD models have proven effectiveness in modeling the meson and baryon spectra~%
\cite{deTeramond:2005su,Erlich:2005qh,Branz:2010ub,Gutsche:2011vb,%
Li:2013oda,Brodsky:2014yha,Chen:2021wzj,Zhang:2021itx,Chen:2022goa}, as well as
form factors and decay constants~%
\cite{Brodsky:2006uqa, Kwee:2007nq, Kwee:2007dd, Grigoryan:2007wn,%
Grigoryan:2007vg, Brodsky:2007hb, Abidin:2008hn, Abidin:2008ku, Abidin:2009aj,%
Abidin:2009hr, Brodsky:2014yha, Ballon-Bayona:2017bwk, Gutsche:2017lyu, Abidin:2019xwu,%
Lyubovitskij:2020gjz, Ahmed:2023zkk, Liu:2025vfe}.
In these models, baryons are introduced into the bulk via five-dimensional Dirac spinor fields.
To represent the chiral structure of baryons properly, two bulk spinors are required,
which are dual to the left- and right-handed chiral baryon operators on the boundary~%
\cite{Hong:2006ta, Henningson:1998cd}.
The chiral symmetry breaking is then implemented through the Yukawa coupling between these spinors
and a bifundamental scalar field.
The vacuum expectation value of the scalar field encodes both the explicit and spontaneous symmetry
breaking through its dependence on the quark masses and condensates.
This setup allows one to derive coupled equations of motion in the bulk, whose solutions yield
the baryon wave functions, mass spectra, and form factors.

In holographic QCD, GFFs are computed by perturbing the bulk metric and examining the
response of the baryons to this perturbation.
The fluctuation of the five-dimensional metric is dual to the EMT on the boundary.
The bulk-to-boundary propagator for this fluctuation determines the coupling of the baryon fields
to the external EMT operator.
This construction thus enables the calculation of correlation functions involving the EMT
operator, from which GFFs can be extracted~\cite{Abidin:2008ku}.

In this paper, we apply the bottom-up holographic QCD approach to systematically study
GFFs of the spin-$\frac{1}{2}$ baryon octet.
We include chiral symmetry-breaking terms in the bulk Dirac action, which account
for flavor-dependent effects in the baryon sector.
The equations of motion for the baryon wave functions are then derived, and
adjustable parameters are fixed by the mass eigenvalues.
We then evaluate the GFF $A(Q^2)$ by coupling the baryons to the transverse-traceless
bulk graviton, and obtain the mass radius from the slope of $A(Q^2)$ near $Q^2 \to 0$.

Our results for $A(Q^2)$ in the nucleon sector are in good agreement with those obtained
from lattice QCD calculations.
Furthermore, we find that the gravitational radii of the hyperons decrease with the increasing
strangeness content, reflecting enhanced energy localization in heavier quark sectors.
These findings extend previous holographic studies of GFFs from the nucleon sector
to the entire baryon octet and provide new predictions for mass radii of the hyperons.

The remainder of this paper is organized as follows.
In section~\ref{sec:2}, we introduce our holographic QCD model for spin-$\frac{1}{2}$
baryons, describing the bulk action, scalar field structure, and implementation of
the chiral symmetry breaking.
Section~\ref{sec:3} explains the method for computing GFFs, including the perturbative
graviton coupling and the three-point correlation functions.
In section~\ref{sec:4}, we present the numerical results, including the mass spectra,
gravitational radii, and comparison with lattice QCD and other approaches.
Section~\ref{sec:5} summarizes the present study and discusses possible directions
for future research.
In the appendix, we examine the local behavior of the wave functions,
which can be extracted from their equations of motion.

\section{Baryon octet in the holographic QCD model}
\label{sec:2}

In this section, we present a description of the spin-$\frac{1}{2}$
baryon octet in the bottom-up holographic QCD framework.
This approach is based on the gauge/gravity duality, which states that a strongly
coupled four-dimensional gauge theory, such as low-energy QCD, can be effectively
mapped onto a weakly coupled gravitational theory in a five-dimensional curved
background, typically the AdS space.
Although QCD itself is not conformal, the AdS geometry provides a convenient
starting point.
By introducing hard- or soft-wall cutoffs, one can capture essential
nonperturbative QCD features, including the confinement and spontaneous chiral symmetry breaking.
To incorporate baryonic degrees of freedom, we follow the standard procedure
introduced in refs.~\cite{Hong:2006ta, Henningson:1998cd}, in which two five-dimensional
Dirac spinor fields, $B_k$ $(k=1,2)$, are introduced in the bulk.
These fields are dual to the left- and right-handed baryon operators, $\mathcal{O}_L$ and
$\mathcal{O}_R$, in the field theory on the boundary and transform appropriately under
the flavor symmetry group SU(3).
The doubling of the spinor fields is necessary to correctly implement the parity
structure, chiral-invariant mass terms, and chiral symmetry breaking in the bulk action.

The action is decomposed into two parts:
\begin{equation}
S = S_{0} + S_{\chi},
\end{equation}
where $S_{0}$ is the chiral-invariant term containing the kinetic and
covariant interaction terms, while $S_{\chi}$ encodes the couplings
responsible for the chiral symmetry breaking via the scalar field background.
The explicit form of the chiral-invariant term is
\begin{equation}
 S_{0} = \int d^5 x \, \sqrt{g} \sum_{k=1}^2 \text{Tr} \Big[ \bar{B}_k \, i \,
  e_A^M \Gamma^A \nabla_M B_k - m_5 \bar{B}_k B_k \Big],
\end{equation}
where $B_1$ and $B_2$ are bulk baryon spinors, $m_5$ is the 5D conformal mass,
$g = |\det g_{MN}|$ is the determinant of the background AdS metric $g_{MN}$, $e_A^M$ is
the vielbein defined via $g_{MN} =e_M^A e_N^B \eta_{AB}$, and $\Gamma^A = (\gamma^\mu, -i\gamma^5)$
are the Dirac gamma matrices satisfying $\{ \Gamma^A, \Gamma^B \} = 2\eta ^{AB}$.
We adopt the standard Poincar\'e patch as the coordinate system for AdS$_5$:
\begin{equation}
 ds^2 = \frac{1}{z^2} \left( \eta_{\mu\nu} dx^\mu dx^\nu - dz^2 \right).
\end{equation}
We introduce a UV cutoff $\epsilon \to 0$ and an IR cutoff $z_0 \sim 1/\Lambda_{\text{QCD}}$
to represent the confinement scale. 
Accordingly, the coordinate $z$ is restricted to the domain $z \in [\epsilon, z_0]$.
The covariant derivative acting on the Dirac spinors includes both spin and gauge connections:
\begin{equation}
\nabla_M = \partial_M + \frac{i}{4} \omega_M^{AB} \Gamma_{AB} - i A_{Ma}^{L(R)} t^a,
\end{equation}
where $\omega_M^{AB}$ is the spin connection, $\Gamma_{AB} = \frac{1}{2i}[\Gamma^A, \Gamma^B]$
are the generators of local Lorentz transformations, $t^a$ are the SU(3) flavor generators,
and $A_M^{L(R)}$ are the bulk gauge fields dual to the chiral $\mathrm{SU(3)}_{L,R}$ currents
on the boundary. 
For the AdS$_5$ background, the nonvanishing components of the spin connection are:
\begin{equation}
\omega_\mu^{z\nu} = - \omega_\mu^{\nu z} = \frac{1}{z} \delta^{\nu}_\mu.
\end{equation}

The baryon fields are arranged into SU(3) flavor-octet matrices as
\begin{equation}
 B_{1,2} = \begin{pmatrix}
\frac{\Sigma^0}{\sqrt{2}} + \frac{\Lambda}{\sqrt{6}} & \Sigma^+ & p \\
\Sigma^- & -\frac{\Sigma^0}{\sqrt{2}} + \frac{\Lambda}{\sqrt{6}} & n \\
\Xi^- & \Xi^0 & -\frac{2\Lambda}{\sqrt{6}}
    \end{pmatrix}_{L,R}.
\end{equation}
To incorporate the chiral symmetry breaking, we introduce a scalar field
$X$ which transforms in the bifundamental representation of $\mathrm{SU(3)}_L \times \mathrm{SU(3)}_R$
and acquires a non-trivial vacuum expectation value $X_0(z)$.
The Yukawa-type coupling between the baryon fields and $X$ is responsible for the baryon mass generation
and parity mixing.
The symmetry-breaking part of the action is given by
\begin{equation}
\begin{aligned}
S_{\chi} = - \int d^5x \sqrt{g} \big[ & c_1\, \text{Tr}(\bar{B}_1 \{ \chi_+, B_2 \}) + c_2\, \text{Tr}(\bar{B}_1 [\chi_+, B_2]) \\
& + (c_2 - c_1)\, \text{Tr}(\bar{B}_1 B_2) \text{Tr}(\chi_+) + \text{h.c.} \big],
\end{aligned}
\end{equation}
where $\chi_+ = \frac{1}{2}(\xi^\dagger X \xi^\dagger + \xi X^\dagger \xi)$ and
$\xi = \exp(i t^a \pi_a)$ with $\{\pi_a\}$ denoting the bulk pseudoscalar meson fields.
The scalar field $X$ is expressed as $X = \xi X_0 \xi$, where
\begin{equation}
X_0(z) = \frac{1}{2} \begin{pmatrix}
v_u(z) & 0 & 0 \\
0 & v_d(z) & 0 \\
0 & 0 & v_s(z)
\end{pmatrix} \quad \text{with} \quad v_i(z) = m_i z + \sigma_i z^3.
\end{equation}
The index $i$ runs over the flavors $u, d$, and $s$, with $m_i$ and $\sigma_i$ corresponding
respectively to the current quark mass and the chiral condensate for flavor $i$.

For the extraction of the physical spectra and wave functions, the five-dimensional (5D)
bulk fields are decomposed into the four-dimensional (4D) Kaluza-Klein (KK) modes as follows:
\begin{align}
\begin{pmatrix} B_{1L} \\ B_{2L} \end{pmatrix} &= \sum_n \begin{pmatrix} f_{1L}^{(n)}(z)\, \psi_L^{(n)}(x) \\ f_{2L}^{(n)}(z)\, \psi_L^{(n)}(x) \end{pmatrix}, \\
\begin{pmatrix} B_{1R} \\ B_{2R} \end{pmatrix} &= \sum_n \begin{pmatrix} f_{1R}^{(n)}(z)\, \psi_R^{(n)}(x) \\ f_{2R}^{(n)}(z)\, \psi_R^{(n)}(x) \end{pmatrix}.
\label{KK}
\end{align}
Here $\psi^{(n)}(x)$ are 4D spinor fields corresponding to the $n$-th KK mode, while
the KK wave functions $f_{kL}$ and $f_{kR}$ describe the profiles of the bulk fields along the
extra dimension $z$.
These profiles encode the wave-function structure of each mode in the fifth dimension.
The KK wave functions are normalized according to the orthonormality conditions
\begin{equation}
\int_0^{z_0} \frac{dz}{z^4} \left( |f_{1L}^{(n)}|^2 + |f_{1R}^{(n)}|^2 \right) = \int_0^{z_0} \frac{dz}{z^4} \left( |f_{2L}^{(n)}|^2 + |f_{2R}^{(n)}|^2 \right) = 1.
\end{equation}
This normalization ensures that the 4D fields $\psi^{(n)}(x)$ have canonical kinetic terms
in the effective 4D theory, allowing for a direct identification of the physical spectra of baryons
and their corresponding wave functions.
The variation of the total action leads to coupled equations of motion for the baryon fields.
Decomposing the 5D spinors into left- and right-handed chiral components, one obtains
\begin{align}
\begin{pmatrix}
\partial_z - \frac{2 + m_5}{z} & -\frac{v(z)}{2z} \\
-\frac{v(z)}{2z} & \partial_z - \frac{2 - m_5}{z}
\end{pmatrix}
\begin{pmatrix}
f_{1L}^{(n)}\\ f_{2L}^{(n)}
\end{pmatrix}
&= -|p|\begin{pmatrix} f_{1R}^{(n)} \\ f_{2R}^{(n)} \end{pmatrix},\label{EOM1}\\
\begin{pmatrix}
\partial_z - \frac{2 - m_5}{z} & \frac{v(z)}{2z} \\
\frac{v(z)}{2z} & \partial_z - \frac{2 + m_5}{z}
\end{pmatrix}
\begin{pmatrix}
f_{1R}^{(n)}\\ f_{2R}^{(n)}
\end{pmatrix}
&= +|p|\begin{pmatrix} f_{1L}^{(n)} \\ f_{2L}^{(n)} \end{pmatrix},\label{EOM2}
\end{align}
where $v(z)$ serves as the flavor-dependent effective scalar coupling.
Its form depends on the baryon species and is explicitly given by
\begin{equation}
v(z) = 
\begin{cases}
(3c_2 - c_1)\, v_u(z), & \text{for nucleons } (p,n), \\
\left(2c_2 - \frac{4c_1}{3} \right) v_u(z) + \left( \frac{c_1}{3} + c_2 \right) v_s(z), & \text{for } \Lambda, \\
2c_2\, v_u(z) + (c_2 - c_1) v_s(z), & \text{for } \Sigma, \\
(c_2 - c_1) v_u(z) + 2c_2\, v_s(z), & \text{for } \Xi.
\end{cases}
\end{equation}
The boundary conditions are imposed to ensure both proper normalization and
consistency with the required parity properties:
\begin{equation}
f_{1L}^{(n)}(\epsilon) = 0, \qquad f_{1R}^{(n)}(z_0) = 0.
\end{equation}
The parity-even and parity-odd states satisfy
\begin{align}
\text{Even parity:} \quad & f_{1L}^{(n)} = +f_{2R}^{(n)}, \quad f_{1R}^{(n)} = -f_{2L}^{(n)}, \label{even}\\
\text{Odd parity:} \quad & f_{1L}^{(n)} = -f_{2R}^{(n)}, \quad f_{1R}^{(n)} = +f_{2L}^{(n)}.\label{odd}
\end{align}
The KK wave functions $f_{L,R}^{(n)}(z)$ determine not only the baryon mass spectra but also
their couplings to external currents, and consequently the form factors.
They play a central role in the subsequent evaluation of GFFs in the next section.

\section{Gravitational form factors}
\label{sec:3}

Beyond electromagnetic probes, GFFs provide a complementary and theoretically profound
means of exploring the internal structure of hadrons.
They are defined through matrix elements of the EMT $T^{\mu\nu}$, a conserved Noether current
associated with spacetime translations.
Although direct measurement of gravitational couplings is not experimentally feasible at
the hadronic scale, GFFs can be accessed indirectly from GPDs through sum rules applied
to deeply virtual exclusive processes.

In the context of spin-$\frac{1}{2}$ baryons, the matrix element of the symmetric EMT
between one-particle states is parameterized by three independent GFFs:
\begin{align}
&\langle p_2, s_2 | T^{\mu\nu}(0) | p_1, s_1 \rangle 
\notag\\
&= \bar{u}(p_2, s_2) \left[ A(Q^2) \gamma^{(\mu} P^{\nu)} + B(Q^2) \frac{i P^{(\mu} \sigma^{\nu)\alpha} q_\alpha}{2m} + D(Q^2) \frac{q^\mu q^\nu - q^2 \eta^{\mu\nu}}{4m} \right] u(p_1, s_1),
\end{align}
where $u(p, s)$ is the positive energy Dirac spinor, $q = p_2 - p_1$ the momentum transfer,
$P = \frac{1}{2}(p_1 + p_2)$ the average momentum, and $m$ the baryon mass.
Parentheses around indices denote symmetrization.
The GFF $A(Q^2)$ is associated with the momentum fraction carried by constituents
and satisfies the normalization condition $A(0) = 1$ due to the momentum conservation.
$B(Q^2)$ encodes contributions of quarks and gluons to the spin structure of the baryon.
$D(Q^2)$ governs the spatial distribution of internal stress, including the pressure and the shear forces.

In the holographic framework, these GFFs arise from the interaction of the baryon fields with
fluctuations of the bulk metric $g_{MN}$, which are dual to insertions of the EMT operator
$T^{\mu\nu}$ in the field theory on the boundary.
The metric is perturbed as
\begin{equation}
g_{\mu\nu}(x,z) = \eta_{\mu\nu} + h_{\mu\nu}(x,z),
\end{equation}
where $h_{\mu\nu}$ is the symmetric fluctuation field in AdS$_5$ corresponding
to the external source for $T^{\mu\nu}$.
Under the transverse-traceless gauge conditions, $\partial^\mu h_{\mu\nu} = 0$ and $h^\mu_\mu = 0$,
the graviton fluctuation can be factorized as
\begin{equation}
h_{\mu\nu}(q,z) = \mathcal{H}(q,z) h_{\mu\nu}^0(q),
\label{factorization}
\end{equation}
where $h_{\mu\nu}^0(q)$ denotes the Fourier-transformed source field on the boundary,
and $\mathcal{H}(q,z)$ is the bulk-to-boundary propagator.
The equation of motion for $\mathcal{H}$ reads
\begin{equation}
\left( \partial_z \frac{\partial_z}{z^3} + \frac{1}{z^3}q^2\right)\mathcal{H}(q,z) = 0.
\end{equation}
For the space-like momentum transfer $q^2 = -Q^2 < 0$, the graviton propagator in
the hard-wall AdS/QCD model takes the form~\cite{Abidin:2008ku}:
\begin{equation}
\mathcal{H}(Q,z) = \frac{Q^2 z^2}{2} \left( \frac{K_1(Q z_0)}{I_1(Q z_0)} I_2(Q z) + K_2(Q z) \right),
\end{equation}
where $I_n$ and $K_n$ are the modified Bessel functions of the first and second kind, respectively.

In order to extract the baryon response to the gravitational source, we examine
the variation of the action under the metric perturbation.
The kinetic term $S_{0}$ induces a coupling between the baryon fields and $h_{\mu\nu}$ of the form:
\begin{equation}
S_{\text{int}} = \int d^5x\, \frac{-i}{4z^4}\, h_{\mu\nu}(x,z)\, \bar{B}_1(x,z)\, \gamma^\mu\, \partial^\nu B_1(x,z).
\end{equation}
This vertex represents the baryon-graviton interaction in the bulk.
After Fourier transforming the baryon fields and inserting the factorized graviton,
eq.~\eqref{factorization}, the variation of the action becomes
\begin{align}
S_{\text{int}} &= \int \frac{dz}{z^4} \int \frac{d^4p_2\, d^4q\, d^4p_1}{(2\pi)^{12}}\, (2\pi)^4 \delta^4(p_2 - q - p_1)\, h_{\mu\nu}^0(q)\, \mathcal{H}(q,z) \notag \\
&\quad \times \left[ \bar{B}_{1L}(p_2,z)\, \gamma^\mu\, p^\nu\, B_{1L}(p_1,z) + \bar{B}_{1R}(p_2,z)\, \gamma^\mu\, p^\nu\, B_{1R}(p_1,z) \right].
\end{align}
Using the KK decomposition \eqref{KK} for the 5D baryon fields, and projecting onto on-shell 4D
spinors $\psi_{L,R}^{(n)}(x)$, the form factor $A(Q^2)$ can be read off from the overlap integrals
of the bulk baryon wave functions with the graviton propagator.
The resulting form factor is
\begin{equation}
A(Q^2) = \int_{\epsilon}^{z_0} \frac{dz}{2z^4}\, \mathcal{H}(Q,z) \left[ \left| f_{1L}^{(n)} (z) \right| ^2 + \left| f_{1R}^{(n)} (z) \right| ^2 \right]. 
\end{equation}
Owing to the parity relations between $B_1$ and $B_2$, an equivalent expression in terms of $f_{2L}$
and $f_{2R}$ also holds.
The gravitational radius is then obtained from the slope of $A(Q^2)$ at zero momentum transfer:
\begin{equation}
\langle r_G^2 \rangle = -6 \left. \frac{dA(Q^2)}{dQ^2} \right|_{Q^2 = 0}.
\end{equation}

\section{Numerical results}
\label{sec:4}

In this section, we present a detailed numerical analysis of the GFF $A(Q^2)$ for the spin-$\frac{1}{2}$
baryon octet, computed in the bottom-up holographic QCD framework developed in sections~\ref{sec:2}
and \ref{sec:3}.
A distinctive feature of this approach is that the GFFs are derived without introducing any additional
free parameters beyond those already fixed by fitting the empirical baryon mass spectra.
Consequently, once the model parameters are determined, the resulting GFFs and the corresponding mass radii
constitute genuine predictions of our holographic model.
Under the specific choice $m_{5}=3/2$, we fix the seven parameters $\{c_1,c_2,\sigma_q,\sigma_s,m_q,m_s,z_0\}$
through a global fit to the ground-state masses of the baryon octet.
For each trial parameter set, the coupled five-dimensional Dirac equations (eqs.~\eqref{EOM1} and \eqref{EOM2})
are solved in every flavor channel, using the standard UV/IR hard-wall boundary conditions.
The lowest normalizable eigenvalue in each channel is then taken as the output of the model,
identified with the corresponding baryon masses.
We compare these outputs with the empirical baryon masses reported by the Particle Data Group
(PDG)~\cite{ParticleDataGroup:2024cfk}, and the parameters are determined by minimizing a weighted $\chi^2$
function subject to the positivity constraints, $\sigma_{q,s}>0$, $m_{s}>m_{q}>0$, and $z_0>0$.

The resulting best-fit parameter values are summarized as follows:
\begin{equation}
\begin{aligned}
c_1 &= -0.8137 \pm 0.035\,, & c_2 &= 2.740 \pm 0.004 \,, \\[4pt]
\sigma_q &= (301.86 \pm 32~\mathrm{MeV})^3\,, & \sigma_s &= (318.7\pm0.5~\mathrm{MeV})^3\,, \\[4pt]
m_q &= 14.8\pm0.5~\mathrm{MeV}\,, & m_s &= 152.7\pm0.9~\mathrm{MeV}\,, \\[4pt]
z_{0} &= 3.424\pm0.005~\mathrm{GeV}^{-1}\,.
\end{aligned}
\end{equation}
These values are consistent with phenomenological expectations and with previous AdS/QCD
implementations~\cite{Erlich:2005qh, Hong:2006ta}.
The enhancement of $\sigma_s$ relative to $\sigma_q$ indicates a stronger contribution of
the heavier quark sector in the spontaneous chiral symmetry breaking, whereas the IR cutoff $z_0$
corresponds to a confinement scale of $\Lambda_{\mathrm{QCD}}\sim 290~\mathrm{MeV}$.

\begin{figure}[tb]
  \centering
  \includegraphics[width=0.9\textwidth]{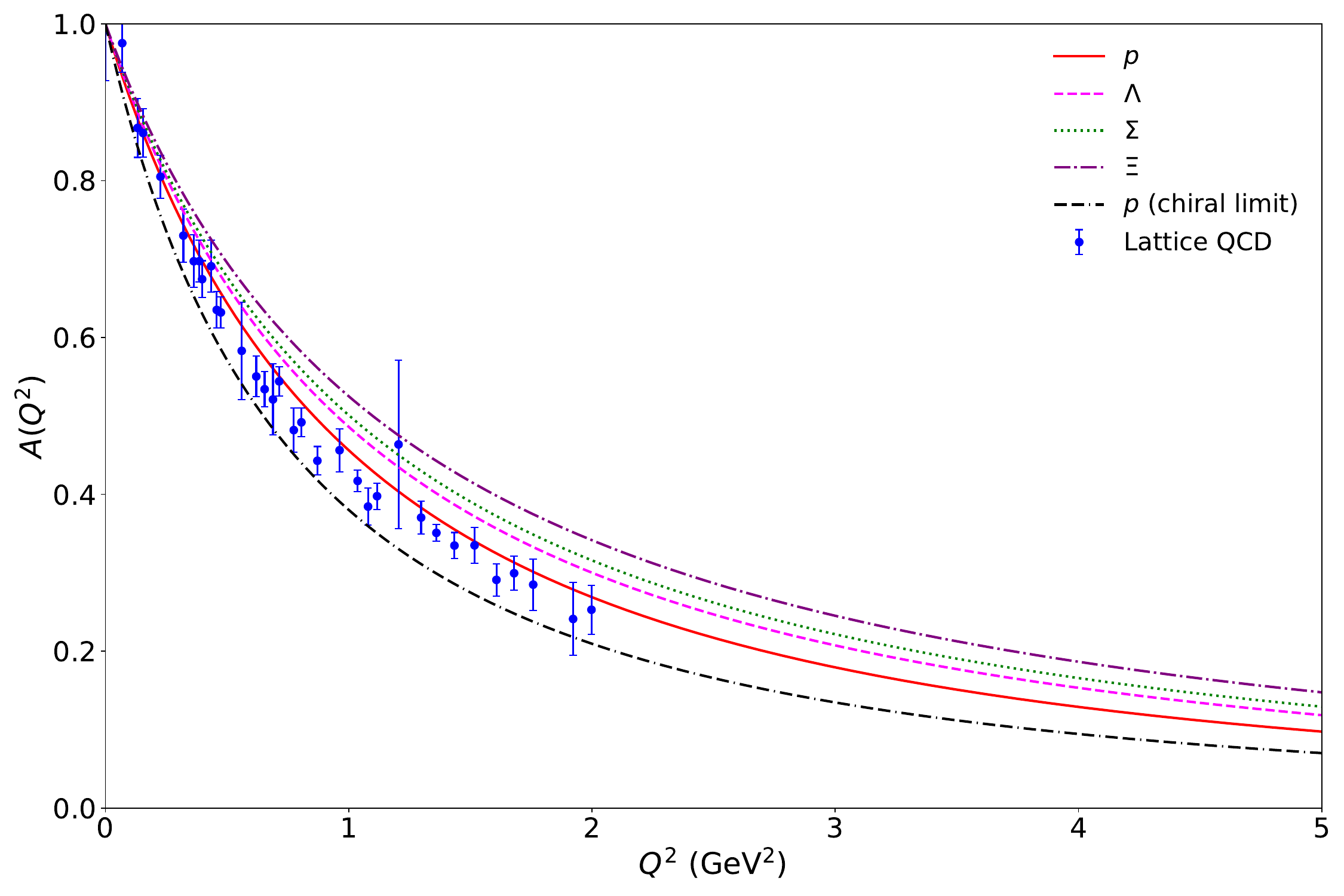}
 \caption{The GFF $A(Q^2)$ of the baryon octet as a function of $Q^2$.
 Data points with error bars represent lattice QCD results~\cite{Hackett:2023rif} for the proton.
 The model curves are shown as red solid for $p$, dashed for $\Lambda$, dotted for $\Sigma$,
 and dash‐dotted for $\Xi$.
The black double dash-dotted line denotes the prediction for the proton in the chiral limit obtained with a holographic QCD model~\cite{Abidin:2008hn}.}
  \label{fig:AQ2_plot}
\end{figure}

Figure~\ref{fig:AQ2_plot} displays the resulting $A(Q^2)$ for representative members of the baryon octet:
$p$, $\Lambda$, $\Sigma$, and $\Xi$.   
$A(Q^2)$ was also computed in an earlier study using the bottom-up holographic QCD~\cite{Abidin:2008hn},
as shown by the black curve in the figure, which lies slightly below the lattice QCD results.
This discrepancy may arrise from the use of the chiral limit in those calculations,
where the light-quark masses are set to zero and the chiral symmetry remains exact.
In such a treatment, the baryon masses and energy-momentum distributions are shifted,
leading to a systematic suppression of $A(Q^2)$. 
By contrast, our framework incorporates the explicit chiral symmetry breaking through non-zero
quark masses and the chiral condensates, providing a more realistic description of baryon
structure and yielding closer agreement with lattice QCD results.
For all members of the baryon octet, $A(Q^2)$ exhibits a smooth fall-off with increasing momentum transfer, a behavior typical of the Fourier transform of localized spatial energy densities.
The inclusion of strange quarks systematically shifts the curves toward a slower fall-off,
indicating a more compact spatial localization of energy density in hyperons.
The extracted radii are listed in Table~\ref{tab:RGtable}.
\begin{table}[tb]
\centering
\caption{Gravitational RMS radii $\sqrt{\langle r_G^2 \rangle}$ for the baryon octet.}
\label{tab:RGtable}
\begin{tabular}{lc}
\hline
\textbf{Baryon} & $\sqrt{\langle r_G^2 \rangle}$ [fm] \\
\hline
$p$           & 0.488 \\
$\Lambda $     & 0.470 \\
$\Sigma$      & 0.461 \\
$\Xi$       & 0.446 \\
\hline
\end{tabular}
\end{table}
A clear ordering is observed:
\begin{gather}
\sqrt{\langle r_G^2 \rangle}_{p} > \sqrt{\langle r_G^2 \rangle}_\Lambda > \sqrt{\langle r_G^2 \rangle}_\Sigma > \sqrt{\langle r_G^2 \rangle}_\Xi,
\end{gather}
indicating that the mass radii decrease as the strangeness increases in the baryon octet.
In a constituent-quark picture, this tendency can be attributed to the larger constituent
mass of the strange quark, which shifts the center of gluonic momentum to shorter distances.
\section{Conclusion}
\label{sec:5}

We have investigated the GFF $A(Q^2)$ of the spin-$\tfrac12$ baryon octet in a bottom-up 
holographic QCD model with the hard IR wall.
Since the all model parameters are determined by the octet ground-state masses, the resulting $A(Q^2)$ and mass radii are thus genuine predictions.

Effects of the both explicit and dynamical chiral symmetry breaking are taken into account 
by introducing the Yukawa sector $S_\chi$, in which the bilinear field $X_0(z)$ encodes 
the current quark masses and chiral condensates. 
The seven model parameters are determined solely from the octet masses; $A(Q^2)$ is then 
predicted without any further adjustment. 
Compared with the chiral limit case, inclusion of the explicit chiral symmetry breaking modifies 
the bulk baryon wave functions and, consequently, the overlap with the bulk graviton. 
This leads to systematic changes in the low-$Q^2$ behavior of $A(Q^2)$ that improve agreement
with the lattice results for the proton, while involving no additional free parameters.

Beyond the nucleon, the SU(3) flavor symmetry breaking is introduced through different 
inputs for the light and strange quarks in $X_0(z)$, namely, the flavor dependence 
of $v(z)$ in the wave function.
This modifies the hyperon profiles and their overlap with the graviton mode. 
In the momentum space, $A(Q^2)$ exhibits a slower fall-off as the
strange quark content increases, which manifests as smaller gravitational radii for the hyperons 
in coordinate space.
Within the octet, the nucleon has the largest radius, hyperons with one strange quark are 
smaller, and those with two strange quarks are the smallest.

This work extends previous holographic studies from nucleons to the full baryon octet by 
including the strange quark mass and condensate, with a particular focus on the GFF $A$-term.
An extension of the current work to investigate the GFF $D$-term will deserve detailed future studies, as it can reveal the pressure and shear force distributions within 
hadrons and plays a critical role in their mechanical stability.
It is also worth extending the analysis to spin-$\frac{3}{2}$ baryons and exploring the 
gluon contributions to the EMT. 
Pursuing these directions can provide deeper and more comprehensive insights into 
how masses and forces are distributed within hadrons, highlighting the rich internal 
structure of baryons.

\acknowledgments
The work of Z.L. was supported by the THERS Make New Standards Program for the Next Generation Researchers at Nagoya University (No. 2562409026).
The work of A.W. was supported by JSPS KAKENHI Grant Number JP25K23385.

\appendix
\section{Local analysis for wave equations}
Here we investigate the differential equations, eqs.~\eqref{EOM1} and \eqref{EOM2}, for the
KK wave functions $f_{kL}$ and $f_{kR}$ $(k=1,2)$. 
Although, to our knowledge, exact solutions are not available, the local
behavior of the wave functions can still be extracted.

We first examine the small-$z$ region, which corresponds to the UV limit of the theory.
In this limit, the terms proportional to the quark mass $m_q$ and $p$ are subleading,
and the $\sigma$-dependent terms give next-to-subleading contributions.
Combining eqs.~\eqref{EOM1} and \eqref{EOM2} and dropping $\sigma$-dependent terms,
we obtain four decoupled second-order differential equations of Bessel-type.
The equations for $\{f_{1L},f_{2R}\}$ share the same structure, whereas those
for $\{f_{1R},f_{2L}\}$ take the same form with the sign of $m_5$ reversed.
Factoring out $z^{\frac{5}{2}}$ reduces each of these equations to the standard
Bessel equation,
and the two independent solutions are therefore 
$z^{\frac{5}{2}}J_{\left|m_5\mp\frac{1}{2}\right|}\Big(\sqrt{p^2-\frac{m_q^2}{4}}z\Big)$ and $z^{\frac{5}{2}}Y_{\left|m_5\mp\frac{1}{2}\right|}\Big(\sqrt{p^2-\frac{m_q^2}{4}}z\Big)$,
where the upper sign corresponds to $\{f_{1L},f_{2R}\}$ and the lower sign to
$\{f_{1R},f_{2L}\}$.
Expanding these solutions further in $z$, and retaining only their leading powers in $z$,
--- i.e., dropping $p$- and $m_q$-dependent subleading pieces ---,
yields the pair of leading behaviors $\{z^{2\pm{m_5}},z^{3\mp{m_5}}\}$.
The general solutions to leading order can then be easily constructed as
\begin{align}
 f_{1L}&\sim C_{1}(1+2m_{5})z^{2+m_{5}}%
 +\left(C_{2}|p|+\frac{m_q}{2}D_2\right)z^{3-m_{5}},
 \label{smallz1}\\
 f_{2R}&\sim D_{1}(1+2m_{5})z^{2+m_{5}}%
 +\left(D_2|p|+\frac{m_{q}}{2}C_2\right)z^{3-m_{5}},%
\label{smallz2}\\
 f_{1R}&\sim\left(C_1|p|-\frac{m_{q}}{2}D_1\right)z^{3+m_{5}}%
 -C_{2}(1-2m_{5})z^{2-m_{5}},
\label{smallz3}\\
 f_{2L}&\sim-\left(D_{1}|p|-\frac{m_q}{2}C_1\right)z^{3+m_{5}}%
 +D_{2}(1-2m_{5})z^{2-m_{5}}, 
\label{smallz4}
\end{align}
where $C_{1}$, $C_{2}$, $D_{1}$, and $D_{2}$ are all integration constants.
In the chiral limit $m_q \to 0$, the mixing between the two
$(k=1,2)$ sectors disappears, and eqs.~\eqref{smallz1} and \eqref{smallz3}
reduce to the expressions given in ref.~\cite{Hong:2006ta}.
The even-parity condition \eqref{even} imposes  $D_1=C_1$ and $D_2=C_2$,
while the odd-parity condition \eqref{odd} gives $D_1=-C_1$ and $D_2=-C_2$.
Since we take $m_{5}=3/2$, the second term in $f_{1R}$ behaves as $z^{1/2}$,
which is non-normalizable. 
Thus, we must set $C_{2}=0$, and hence $D_{2}=0$, irrespective of the parity
assignment. The UV behavior of the wave functions is therefore
\begin{gather}
f_{1L}\sim z^{7/2}, \quad f_{1R}\sim z^{9/2}, 
\end{gather} 
which has been confirmed to be in agreement with the numerical results in
section~\ref{sec:4}.

Next we consider the large-$z$ behavior.
Although this region lies formally beyond the IR cutoff $z_{0}$,
the behavior around $z\sim z_{0}$ may still be estimated by extrapolating from $z\sim\infty$. 
In this regime, the contribution from $v(z)=m_q z + \sigma z^{3}$, --- in particular, the part arising
from the term proportional to the chiral condensate $\sigma$ --- becomes dominant, because it has
the highest power in $z$.
The wave equation can then be approximated as 
\begin{gather}
\left(\partial_{z}^{2}-\frac{1}{4}\sigma^{2}z^{4}\right)f_{1L,1R}\sim 0,
\label{largez1}
\end{gather}
which has the solution
\begin{gather}
f_{1L,1R}\sim C_{+}\exp\left(\frac{\sigma}{6}z^{3}\right)+C_{-}\exp\left(-\frac{\sigma}{6}z^{3}\right).
\label{largez2}
\end{gather}
Here, $C_{+}$ and $C_{-}$ are constants.
We have confirmed that eq.~\eqref{largez2} reasonably well reproduces the behavior around $z\sim z_{0}$
observed in the numerical computation.
Note that eq.~\eqref{largez2} is also consistent with the previous works based on the chiral
limit \cite{Abidin:2008hn, Kim:2009bp}, in which the exact wave functions were found to be expressed
in terms of the modified Bessel functions.


\bibliographystyle{JHEP}
\bibliography{references}

\end{document}